\documentclass[aps,prl,twocolumn,reprint,superscriptaddress,amsmath,amssymb]{revtex4-1} 

\usepackage{soul} 
\usepackage{graphicx} 
\usepackage{dcolumn}  
\usepackage{bm}       
\usepackage[colorlinks,urlcolor=blue,citecolor=blue,linkcolor=blue,pdfstartview=FitH]{hyperref}
\usepackage{enumerate}
\usepackage{amsmath}
\usepackage{color}
\usepackage{cases}
\usepackage{microtype}
\usepackage{siunitx}
\usepackage{subdepth}
\usepackage{todonotes}
\usepackage[utf8]{inputenc}
\usepackage[T1]{fontenc}
\usepackage{ dsfont }
\usepackage{txfonts} 
\usepackage{tensor}

\colorlet{outcolor}{red!20}
\colorlet{addcolor}{cyan!20}
\colorlet{hlcolor}{yellow!100}

\begin{document}

\title{Vortex spin in a superfluid}

\author{Emil Génetay Johansen}
\affiliation{Optical Sciences Centre, Swinburne University of Technology, Melbourne 3122, Australia}

\author{Tapio Simula}
\affiliation{Optical Sciences Centre, Swinburne University of Technology, Melbourne 3122, Australia}

\begin{abstract}
General relativity predicts that the curvature of spacetime induces spin rotations on a  parallel transported particle. We deploy Unruh's analogue gravity picture and consider a quantised vortex embedded in a two-dimensional superfluid Bose--Einstein condensate. We show that such a vortex behaves dynamically like a charged particle with a spin in a gravitational field. The existence of a vortex spin in a superfluid complements Onsager's prediction of the quantisation of circulation, and is suggestive of potential  quantum technology applications of rotating superfluids. 
\end{abstract}

\maketitle

Vortices have been appreciated as the elemental constituents of hydrodynamic turbulence for a long time and were quantitatively described in the context of classical fluids already by Helmholtz \cite{Helmholtz1867a} and Kirchoff \cite{Kirchhoff1876a}. An observation that a stable vortex structure must belong to a distinct topological class likely inspired Lord Kelvin to put forth his vortex theory of the atom \cite{Kelvin1867a}. During this era, the notion of vorticity also made its appearance in the theory of electromagnetism when Maxwell proposed that the force between magnetic poles would originate from pressure gradients generated by molecular vortices \cite{Maxwell2010a,Falconer2019a}. In 1949 Onsager predicted the existence of quantum mechanical vortices in superfluids noting that the circulation of such a vortex must be an integer multiple of $\kappa=h/m$, where $h$ is Planck's constant and $m$ is the mass of a molecule the fluid is comprised of \cite{onsager_statistical_1949}. This was followed up by Feynman who further developed the idea of quantum turbulence and quantised vortices in the context of superfluid helium \cite{feynman_chapter_1955}. Vortices have since featured prominently in a broad range of quantum systems including Abrikosov \cite{Abrikosov1957a} and Josephson \cite{Josephson1974a} vortices in superconductors, Nielsen--Olesen vortices in high energy physics \cite{Nielsen1973a}, and quantised vortices in superfluid atomic gases \cite{fetter_rotating_2009}.

In a planar superfluid, the interaction between two quantised vortices is mediated by a potential that has the same logarithmic Coulomb gas form as two-dimensional electric charges. This suggests that a quantised vortex may be described as a particle carrying a well defined, topologically protected, charge quantum number $\kappa$. Popov further showed that by viewing vortices as `electrons', the dynamics of the system of vortices satisfies two-dimensional Maxwell's equations of electromagnetic field theory \cite{Popov1973a}. In this case the superfluid flow and the condensate density play the roles of electric and magnetic fields, respectively. 

In addition to the charge $\kappa$, the vortex particle has also been considered to possess a mass \cite{Popov1973a,Thouless2007a,Zwierlein2014a,Simula2018a,Simula2020a}. The electromagnetic corpuscular view of vortices is further corroborated by the observation that when a vortex in a superfluid is transported around a closed loop, each atom located inside the loop contributes a $2\pi$ Aharonov--Bohm-like topological phase such that the atoms may be viewed as flux tubes that the electrically charged vortex encircles \cite{thouless_vortex_1993,Haldane1985a,Polkinghorne2021a}. While the vortex charge $\kappa$ naturally arises as a constant that couples the particle and the field within the electromagnetic picture, interpretation of the vortex mass as a gravitational coupling constant remains less clear. 

Using Witten's Chern--Simons---gravity duality as an inspiration, quasiparticles in planar superfluids were shown to couple topologically to the background fluid, thus enabling gravitational Aharonov--Bohm-like phases \cite{EmilScipost2023a}. Gravity, as demonstrated by Unruh \cite{Unruh1981a,Barcelo2005a}, can be simulated using superfluids \cite{Steinhauer2016a,Eckel2018a,Viermann2022a}. This may be achieved by introducing the square root of the condensate density as a conformal factor for a flat analogue metric, whose gradient corresponds to curvature. In such an analogue picture of gravity, the fluid itself constitutes the universe in which the elementary particles, the vortices (electrons) and the phonons (photons), reside. Consequently, vortices respond to the density gradient of the fluid in a fashion akin to a massive particle subjected to a gravitational field. Moreover, the vortex core itself is a region of varying condensate density suggesting that this curvature could be associated with the gravitational vortex mass.

Considering that a quantised vortex may be viewed as an elementary electrically charged massive particle in an analogue superfluid spacetime, it is then natural to ask if, like an electron, a vortex may also carry a spin quantum number? Here we show, evidenced by direct numerical calculations, that such a vortex spin originates from the internal quasiparticle structure within the vortex core by linking the kelvon quasiparticle amplitudes to a Majorana star-like spin representation. Specifically, we demonstrate that the orientation of such a spin may be brought to a one-to-one correspondence with the physical location of the vortex in a planar superfluid.

\textit{Analogue gravity in a superfluid ---}
Within the superfluid universe picture the acoustic metric, as applied to a Bose--Einstein condensate, is described by the Gross--Pitaevskii equation (GPE) \cite{Barcelo2005a}. Separating the phase $\theta$ and the density $n(\textbf{r},t)=|\psi(\textbf{r},t)|^2$ of the condensate wave function  $\psi(\textbf{r},t)$ via the Madelung transformation $\psi(\textbf{r},t) = |\psi(\textbf{r},t)| e^{i \theta(\textbf{r},t)}$, yields an action $S_{\rm{SU}}$ for this superfluid universe expressed as
\begin{equation}\label{action}
    S_{\rm{SU}} = \frac{\hbar^2}{2m}\int dx^2 dt\left(\left(\nabla \theta\right)^2 + \left(\frac{\nabla|\psi|}{|\psi|}\right)^2\right)|\psi|^2,
\end{equation}
where the terms in the GPE that do not explicitly influence the vortex dynamics have been discarded \cite{groszek_motion_2018,Simula2020a}. We focus on azimuthally symmetric and stationary density profiles with $\varv_i$ denoting the components of the superfluid velocity $\boldsymbol{\varv}_s =(\hbar/m) \nabla \theta$. An important difference between the two terms in Eq.~\eqref{action} is that while the electromagnetic term only yields phases $2 \pi \varw$, where $\varw$ is the integer winding number, the gravitational term may lead to fractional non-abelian phases \cite{EmilScipost2023a}. 
\begin{figure}[!t]
    \center
    \includegraphics[width=\columnwidth]{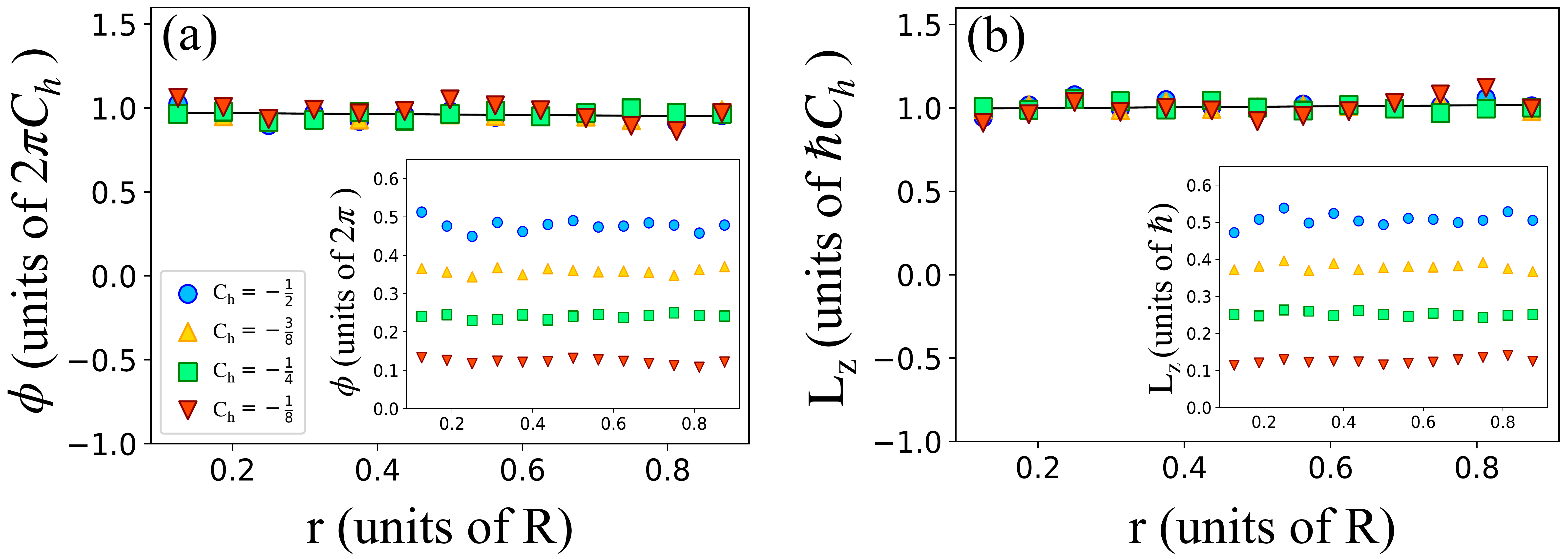}
    \caption{Holonomy transformation. The inset of (a) shows the accumulated geometric phase after the vortex completed a full orbit, inferred from the measurement of the quantum kinetic energy of the state, as functions of the radial vortex position for four different values of $C_h$. The main frame shows the collapse of the data shown in the inset when scaled by the $C_h$ factor. Results in (b) are extracted from the same quantum states as those in (a) with the vortex angular momentum measured using its angular speed.}
    \label{holonomy}
\end{figure}
In Eq.~(\ref{action}) the phase gradient term yields electromagnetism, which is the source of the conventional Aharonov--Bohm-like vortex phase, and the amplitude gradient term yields gravitation in the analogue spacetime \cite{Simula2020a}, which is anticipated to yield a similar topological effect \cite{EmilScipost2023a}. This gravity is captured by the conformal factor $\Omega(r) = |\psi(r)|/|\psi(r_h)|$, of the analogue metric \cite{Barcelo2005a}
\begin{equation}\label{metric}
    g_{\mu \nu}(r) = \Omega^2(r) \left( 
\begin{array}{c|c} 
  -(c^2_s-\varv_s^2) & -\varv_i \\ 
  \hline 
  -\varv_j & \delta_{ij} 
\end{array} 
\right),
\end{equation}
and we define an `elevator speed limit' $c_e=c_s(r_h)=\sqrt{gn(r_h)/m}$ as the speed of sound $c_s$ of the superfluid at the horizon $r_h$. The constant $g$ determines the strength of the  interaction between atoms of mass $m$ within the superfluid, and the horizon $r_h$ is defined as the radial distance from the origin at which the vortex speed $\varv_d$ would equal the local speed of sound.
 
We adopt an Einstein--Cartan picture \cite{2006PPN....37..104Z,Eguchi:1980jx}, which allows for gravity to be formulated as a gauge theory \cite{2018IJGMM..1540005O,10.2307/188767} with the abelian and non-abelian gauge fields $e\indices{^a_\mu}$ and $\omega\indices{^a_{b}}=\omega\indices{^a_{b\mu}}dx^{\mu}$, 
respectively. Here $e\indices{^a_\mu}$ are the tetrad fields defined as the `square root' of the metric $g_{\mu \nu} = e\indices{^a_{\mu}} e\indices{^b_{\nu}} \eta_{ab}$, with $\eta_{ab}$ the Minkowski metric. The gauge field $\omega\indices{^a_{b}}$ is called the spin-connection and relates to the density gradient. Gravitational interactions may be encoded into a gauge covariant derivative $\mathcal{D}_{\mu} = \partial_{\mu} + e\indices{_\mu^a} P_a + \omega\indices{_\mu^{ab}} M_{ab}$, where $P_a$ generates translations and $M_{ab}$ rotations $J_z$ and boosts $K_x$. This object provides instructions for how vectors and spinors residing in the tangent space transform under parallel transport. The field strengths of these gauge fields are respectively given by the torsion and curvature tensors. Combining the tetrad and the spin-connection into a single gauge field 
$  A_{\mu} = e\indices{_{\mu}^a}P_a + \omega\indices{_{\mu}^{ab}}M_{ab}, $ 
the gravity in (2+1)-dimensions can be formulated as a Chern--Simons theory, whose solution is given by the particular configuration $A_{\mu}$ with vanishing field strengths \cite{1988NuPhB.311...46W,ACHUCARRO198689,2009SIGMA...5..080W}. For a vanishing cosmological constant $\Lambda=0$, the generators $M_{ab}$ and $P_a$ are those of the Poincaré group $\rm{ISO}^+(1,2)$. This gauge formalism facilitates a description of the interaction between spin and gravity since the spin resides in the flat tangent space in which the implementation of finite-dimensional representations of the general covariance group are possible, which is not the case globally due to the curvature. When a spinor is parallel transported in such a curved analogue space-time, the spin-connection implements transformations in the Lorentz group. For a stationary and azimuthally symmetric condensate, where the vortices are kept far away from one another, the Lorentz rotations are accounted for by the spatial components
\begin{equation}
\label{spacesol}
    \omega\indices{^r_\theta}(r) = -\omega\indices{^{\theta}_ r}(r) = \left (r \frac{\partial_r \Omega(r)}{\Omega(r)}+1 \right)d\theta,
\end{equation}
and the Lorentz boosts are implemented via the interactions with the temporal components
\begin{equation}
\label{timesol}
    \omega\indices{^t_r}(r) = \omega\indices{^r_ t}(r) = c_e \frac{\partial_r \Omega(r)}{\Omega(r)}dt.
\end{equation}
 All other components vanish under these conditions. These equations represent the components of the first Cartan's structure equation for the metric considered here \cite{EmilScipost2023a}.

 \begin{figure}[!t]
    \center
    \includegraphics[width=\columnwidth]{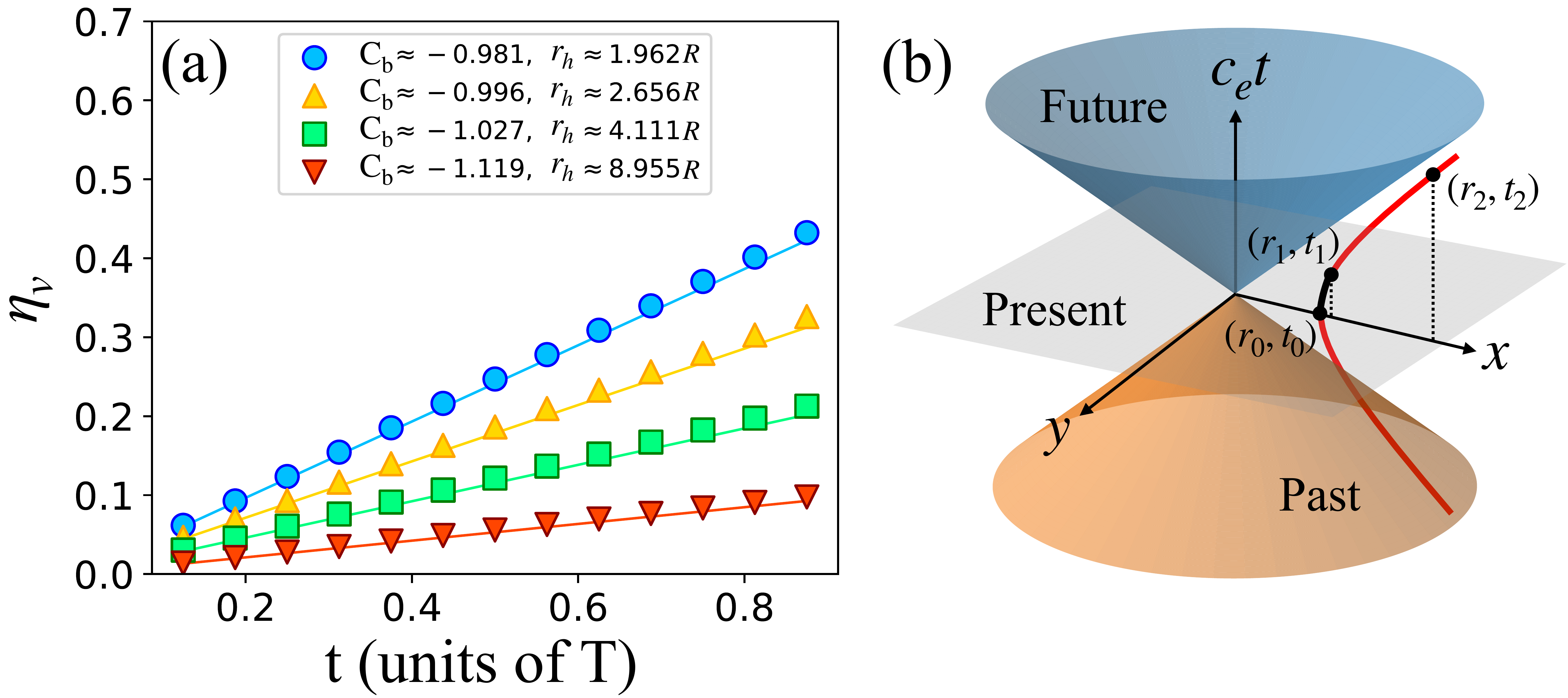}
    \caption{Boost transformation. Frame (a) shows the rapidities $\eta_{\nu}$ according to Eq.~(\ref{spinboost}) as functions of $t$ (solid lines), and the measured distances of the vortex from the gravion $r/r_h = t/t_h$ expressed in terms of the imaginary evolution time $t/T$ (markers), where $T=R / v_d$ and $R$ is the radius of the BEC. The specific $C_b$ and $r_h$ values for each universe is provided in the legend. Frame (b) illustrates a gravion light cone with the hyperbolic vortex world line indicated by the red curve. The line element from $t_0$ to $t_1$ corresponds to the range in which our simulations are carried out. }
    \label{boostB}
\end{figure}

 The transformations (\ref{spacesol}) and (\ref{timesol}) explicitly depend on the conformal factor $\Omega(r)$ meaning that different gauge theories of gravity can be obtained by `spacetime engineering' the condensate density. For instance, any system defined by  a power-law conformal factor $\Omega(r)\propto r^{C_h}$, where $C_h \in \mathds{R}$, is inhabited by a topological defect we refer to as a \emph{gravion}, which upon encirclement enacts a topological non-abelian phase factor $e^{i \pi C_h \sigma_z}$. This is obtained by implementing the spinorial representation of the Lorentz part $\omega\indices{_{\mu}^{ab}}M_{ab}$ of the gauge field $A_{\mu}$, which results in $\omega\indices{^r_{\theta}}\sigma_z/2=C_h\sigma_z/2$. The temporal components vanish at large distance scales for this choice, thus making the spacetime topological. This spacetime thus solves a Chern--Simons gravity. Another interesting case is the exponential conformal factor $\Omega(r)\propto e^{C_br/{r_h}}$, where $C_b\in \mathds{R}$. This results in a constant gravitational field and thus a uniform force everywhere in the system. Since the spatial part of the spin connection vanishes at short distance scales for this $\Omega(r)$, a pure boost may be implemented that is independent of spatial location. 
 
 From the perspective of quantum technology applications these two spacetimes seem particularly interesting since they both benefit from diffeomorphism invariance that protects against external perturbations and noise. Nevertheless, the vortex spin we conjecture exists independent of this choice in all space times. We proceed to simulate vortices embedded in the engineered spacetimes via direct numerical solutions of the Gross--Pitaevskii equation, and attempt to offer an interpretation for `what it is that transforms', i.e. the conjectured vortex spin.

\textit{Lorentz rotations --- }
Let us consider a condensate density profile $n(r)\propto r^{2C_h}$ that yields a conformal factor $\Omega(r)\propto r^{C_h}$. Note that the term
\begin{equation}
    \varv_{d} =\frac{\hbar}{m}\nabla \ln(|\psi|)\cdot\hat{\bf e}_\theta = \frac{\hbar}{m}\frac{\partial_r \Omega(r)}{\Omega(r)},
    \label{veedee}
\end{equation}
equals the speed of the vortex purely due to a condensate density gradient \cite{groszek_vortex_2018}. This allows for an interpretation of
\begin{equation}
    C_h = r \frac{\partial_r \Omega(r)}{\Omega(r)} = \frac{rm{\varv}_{d}}{\hbar} = \frac{L_z}{\hbar},
    \label{CL}
\end{equation}
meaning that the parameter $C_h$ corresponds to an orbital angular momentum of a system where a particle of mass $m$ circles with the orbital speed ${\varv}_{d}$ about the gravion. Attached to each vortex is a quasiparticle, a so-called \emph{kelvon}, which is a zero mode solution to the Bogoliubov--deGennes (BdG) equation in the rotating frame \cite{Polkinghorne2021a}. Transforming to a frame of reference rotating with the angular frequency $L_z/mr^2$, and dropping all the terms that do not explicitly influence the speed of the vortex \cite{groszek_motion_2018}, yields a BdG Hamiltonian
$H=\frac{1}{2m}(-i\hbar\nabla - m{\boldsymbol v}_d)^2 \sigma_z$,
which acts on the kelvon quasiparticle spinor $(u_q,v_q)^T$ that peaks in the vicinity of the vortex core, and is also an exact Nambu-Goldstone zero mode satisfying $|L_z\psi|=|u_q|=|v_q|$\cite{Polkinghorne2021a}. Using Eq.~(\ref{CL}) to express the gauge field as
$m{\boldsymbol v}_d =\hbar{\bf A}_g$
hints at an interpretation of the kelvon as a spin-half object $s=\hbar/2$ minimally coupled to gravity via a gravitational vector potential ${\bf A}_g = C_h \hat{\bf e}_\theta /r$ generated by the topological gravion. The kelvon will consequently transform according to a holonomy that corresponds to a Pauli spin rotation
\begin{equation}\label{spinrot}
    U_h=e^{i \oint \omega\indices{_{\mu}^{ab}}M_{ab} dx^\mu} =e^{i\oint\omega\indices{^r_\theta}(r)\sigma_z/2} =e^{i \oint s{\bf A}_g \cdot d{\boldsymbol\ell}\sigma_z/\hbar}=e^{i \pi C_h \sigma_z}.
\end{equation}
This establishes a map between the vortex motion in physical space determined by $\varv_d$ and a phase rotation of the kelvon spinor determined by the unitary $U_h$, as the action of the time evolution implements a Pauli-$Z$ rotation of the spin. In particular, a $2 \pi$ rotation of the vortex around the gravion in real space leads to a  Pauli-$Z$ rotation of magnitude $2\pi C_h$. The kelvon quasiparticle, which is the true quantum mechanical particle assigned to the vortex, thus acquires a fractional topological phase parametrised by $C_h$ when traversing a closed loop that encloses the gravion in a spacetime defined by $\Omega(r)\propto r^{C_h}$. The vortex, or kelvon, thus forms a `gravion-spin' composite object together with the gravion with a continuous variable mutual exchange phase $\pi C_h$. This is similar to flux attachment of an electron charge and a magnetic flux in fractional quantum Hall fluids with the flux furnished by gravity and charge by the vortex spin. The existence of such a quantum Hall analogy sheds further light on the fact that the gravion solves a Chern--Simons theory, which is known to underpin the quantum Hall effect. In this picture, the parameter $C_h$ thus plays a role similar to the filling fraction.

When the vortex encircles the topological gravion it undergoes a holonomy, which can be deduced by measuring the angular momentum of the vortex. We have demonstrated this by directly simulating the vortex dynamics using the GPE. The result is shown in Fig.~\ref{holonomy}. We measured the total orbital velocity of the vortex $\boldsymbol{\varv}_{\rm tot} = \boldsymbol{\varv}_{s} + \boldsymbol{\varv}_{d}$ by tracking the vortex position in the BEC during real time evolution of the GPE. Following Ref.~\cite{groszek_vortex_2018}, the phase gradient term $\boldsymbol{\varv}_{s}$ was then measured independently and subtracted to yield the value for $\boldsymbol{\varv}_d$. 

In Fig.\ref{holonomy}~(a) the phase $\phi=E_h T_h/\hbar$ scaled by $2 \pi C_h$ is measured for a range of $C_h$ values and vortex distances $r$ from the topological gravion. The measurements of this phase were obtained by calculating the gravitational energy $E_h$ corresponding to the density gradient term in Eq.~\eqref{action} (the quantum pressure), and multiplying it by the measured vortex period $T_h=r/{{\varv}_d}$. Fig.~\ref{holonomy} (a) shows that all geometric phase values collapse on the value $\phi/2\pi C_h = 1$, irrespective of the value of $C_h$ or the vortex radial position. The inset shows the corresponding unscaled raw data. The vortex angular momentum $L_z = mr{\varv}_d$ scaled by $\hbar C_h$ was then measured for the same $C_h$ values and orbital radii as in Fig.~\ref{holonomy} (a). These results, shown in Fig.~\ref{holonomy} (b), are in good agreement with $L_z/\hbar C_h=1$ demonstrating that the gravitational geometric phase can also be inferred from the measurement of the angular momentum of the vortex. The unscaled data are shown in the inset. In combination, the results shown in Figs \ref{holonomy} (a) and (b) constitute independent measurements of the phase $2\pi C_h$ and the vortex angular momentum $L_z$ in Eq.~(\ref{CL}), demonstrating that the azimuthal angle traversed by the vortex in the superfluid directly maps onto the angle $2\pi C_h$ of the Lorentz rotation, Eq.~(\ref{spinrot}) that acts on the kelvon spin.


\textit{Lorentz boosts --- } 
Consider next a pure Lorentz boost generated by a uniform gravity, realised by choosing an exponential conformal factor $\Omega(r)=e^{r{C_b}/{r_h}}$, for which Eq.~(\ref{veedee}) yields 
\begin{equation}
C_b = r_h m{\varv}_{d}/\hbar .
\end{equation}
Geometrically, a boost may also be interpreted as a rotation but in hyperbolic space instead of Euclidean space. This can be understood from the fact that boosts and rotations differ by an imaginary unit $i$. The additional imaginary unit implies that the GPE simulation should be carried out in Wick rotated imaginary time $t \longrightarrow -it$. As the system evolves via the imaginary time propagation, the vortex will, rather than orbiting about the gravion symmetry axis, traverse the system radially towards the (non-vortex ground state) density minimum, or density maximum if the arrow of time is reversed. The radial vortex position thus maps onto the hyperbolic angle and the rapidity of the Lorentz boost. This corresponds to a transformation $ \eta_{\nu} K_x \longrightarrow i \eta_{\nu} \sigma_x/2$, where a hyperbolic rotation about the $x$-direction is generated by $i \sigma_x/2$ with rapidity $\eta_{\nu}$. 
In the BdG description, the spinorial representation of $K_x \longrightarrow i \sigma_x/2$ is implemented but now in imaginary time so that the exponential $\exp{(-iH t/\hbar)}$ becomes $\exp{(-H' t/\hbar)}$, where $H'$ is the Hamiltonian driving the boost. The imaginary time evolution thus enacts a boost transformation  
  \begin{equation}\label{spinboost}
     U_b = e^{\int_0^t\omega\indices{^t_r}(r)\sigma_x/2} = e^{C_b (t/ t_h)  \sigma_x/2}, 
 \end{equation}
where $C_b$ sets the strength of the constant gravity and $t_h=r_h/c_e$ is the boost time when the vortex would reach the horizon. 

We digress to note that the constant $C_b$ parametrises both the quantum pressure, i.e. gravity, and the speed of the vortex. This comes about since in real time propagation a Magnus force $\textbf{F}_M \propto \boldsymbol{\varv}_d \times \hat{\textbf{e}}_z$ is exerted on the vortex. As the vortex travels with a speed $\varv_d$ in real coordinate time, it will also experience an `Einstein elevator' subjected to a constant force $\textbf{F}_M$. The rapidity $\eta_{\nu}=C_b t/t_h$ of the elevator in which the vortex is at rest determines the hyperbolic angle in Eq.~\eqref{spinboost}, where $t$ has the interpretation of a proper time such that the elevator has a constant proper acceleration $a=c_e d \eta_{\nu} /dt = r_h C_b$.
 
Fig.~\ref{boostB} (a) shows the rapidity $\eta_{\nu}$ in Eq.~\eqref{spinboost} as functions of $t$ (solid lines) and the corresponding numerically measured radial positions of the vortex (markers), expressed in temporal units. This result demonstrates that the radial position of the vortex as it evolves in imaginary time equals the rapidity of the elevator as a function of its proper time, such that the speed and position of the vortex map onto the acceleration and speed of the elevator, respectively. Figure~\ref{boostB} (b) shows a light cone centered at the gravion. The section of the hyperbolic curve highlighted in black, from ($r_0,t_0$) to ($r_1,t_1$), represents the range considered in this work. The spacetime point ($r_2,t_2$) corresponds to a radial vortex position in the ultra relativistic regime where the vortex would be traveling close to the local speed of sound. 

\begin{figure}[!t]
    \center
    \includegraphics[width=\columnwidth]{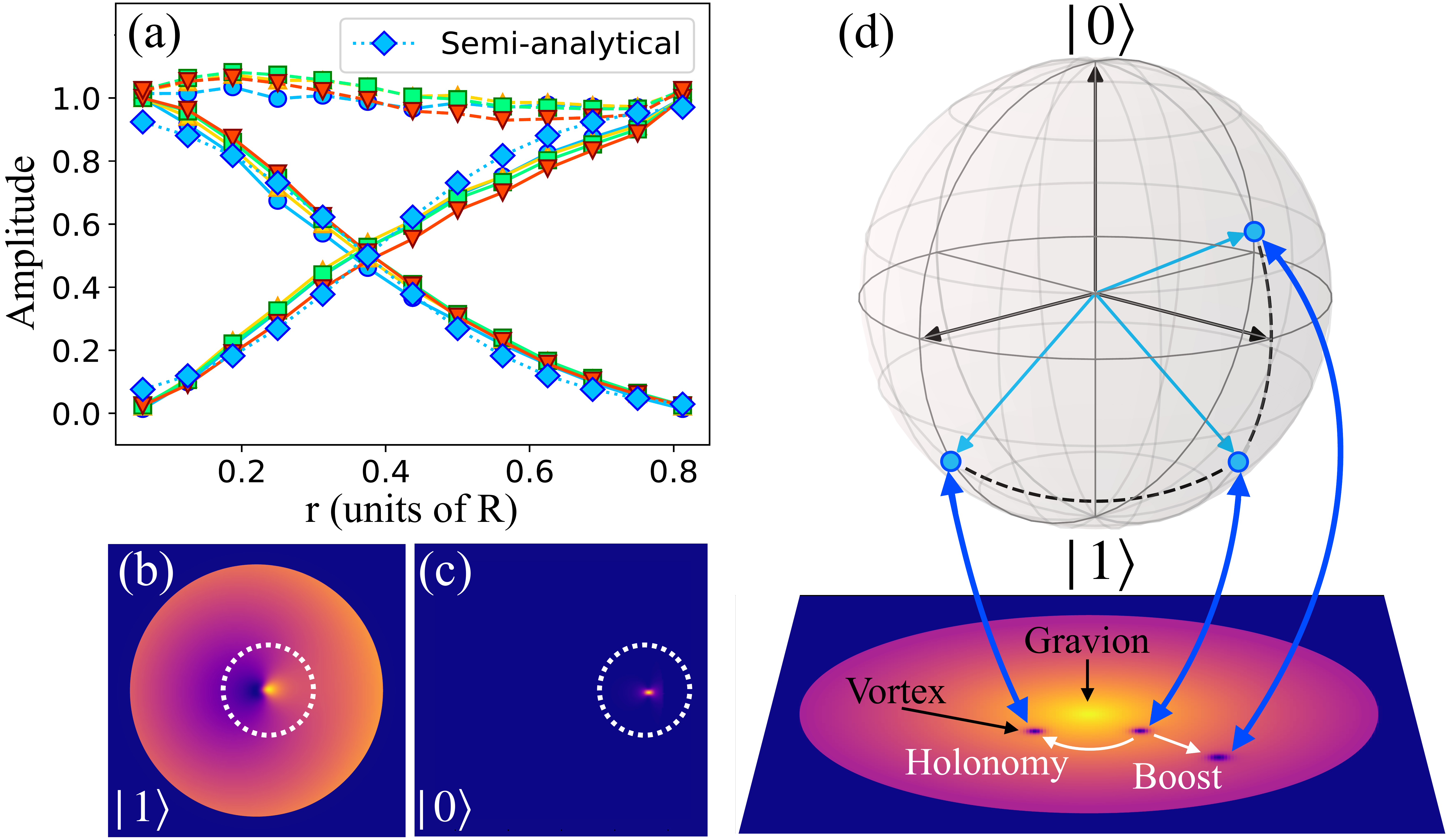}
    \caption{Vortex spin. (a) shows (probability) amplitudes of the kelvon state $\hat{L}_z|\psi\rangle$, projected onto the two basis states $|1\rangle$ and $|0\rangle$ shown in (b) and (c), respectively, and their sum as functions of the boosted radial position of the vortex, measured in units of the numerical system radius $R$. The four marker types correspond to spacetimes defined in the legend of Fig.~2. The diamond markers correspond to spin rotation enacted by Eq.~(\ref{spinboost}) as described in the text. The dashed circles in (b) and (c) show the state projection region, see text, and the kelvon state (orange disk) in (b) has radius $R$. Frame (d) shows the correspondence between the kelvon state vectors on the Bloch sphere and the respective vortex positions in the physical space of the planar Bose--Einstein condensate for $C_b=1$. The radial vortex positions separated by a boost are $r_i =2R/8$ and $r_f =5R/8$, respectively. The kelvon states denoted by circular markers on the Bloch sphere have been extracted as in (a) and normalised thereafter.}
    \label{boostA}
\end{figure}

\emph{Vortex spin---}We have thus numerically demonstrated that the vortex azimuthal real time dynamics realizes Lorentz rotations and that the radial imaginary time dynamics of the vortex realizes Lorentz boost transformations within the analogue gravity picture. Our third key result, highlighted in Fig.~\ref{boostA}, uncovers the spin of the vortex. The kelvon, which is a zero mode resulting from the breaking of $\rm{SO}(2)$ symmetry, has the wave function $\Phi(\textbf{r}) = \hat{{L}}_z \psi(\textbf{r})$ \cite{Polkinghorne2021a}. Figure~\ref{boostA} (b) and (c) show the kelvon densities $|\Phi(r,r_v= 0.5 R/8)|^2$ and $|\Phi(r,r_v=6.5 R/8)|^2$, respectively. We use these two as extremal states to define a near orthonormal basis $|1\rangle$ and $|0\rangle$ for a two-level atom by cropping the kelvon wavefunctions to a region of radius $l_v = 2 \pi \xi$ around the vortex core, indicated by the dashed circles in (b) and (c), with $\xi$ being the vortex healing length. 

Fig.~\ref{boostA} (a) shows the projected kelvon amplitudes $|\langle \Phi|0\rangle|^2$ , $|\langle \Phi|1\rangle|^2$ and their sum (data points near unit amplitude) for four spacetimes as functions of the radial position of the boosted vortex. Also shown is (dotted curve with diamond markers) a semi-analytical result obtained by applying the boost operator of Eq.~\eqref{spinboost} to the equal superposition state $|0\rangle_z=(1,0)^T$ and then evolving forward and backward in time such that $\lim_{t\to\infty}U_b|0\rangle_z\to(1,1)^T/\sqrt{2}=|0\rangle_x$ and $\lim_{t\to-\infty}U_b|0\rangle_z\to(1,-1)^T/\sqrt{2}=|1\rangle_x$. An effective horizon $t/t_h=r/r_h \approx r/2R$ with $C_b=-1$ was used in this calculation and state renormalisation after each boost is actioned since Eq.~\eqref{spinboost} is not unitary.

Figure~\ref{boostA} (d) shows the correspondence between the vortex position in the two-dimensional BEC and the quantum state of its spin-1/2 degree of freedom represented on the Bloch sphere. As illustrated in Fig.~\ref{boostA} (d), there is a one-to-one Majorana star-like map \cite{Majorana1932a} between the vortex position and a Bloch sphere, which furnishes the spin-1/2 degree of freedom of the vortex. The kelvon states in Fig.~\ref{boostA} (b) and (c) bear similarity to the lowest two eigenstates of a two-dimensional quantum harmonic oscillator and offer another interpretation for the vortex spin with the two spin-1/2 eigenstates corresponding to the presence and absence of a quantum depleted atom at the vortex core.
 
We have predicted the existence of a vortex spin in a superfluid, based on Unruh's analogue gravity picture, and provided numerical evidence that the vortex spin arises from the internal kelvon quasiparticle structure of the vortex core and has an analogue gravitational origin. It is amusing to contemplate the reverse analogue whereby the electron spin would originate from an anisotropic probability distribution within the volume of the electron, and be generated by curvature of spacetime.

\begin{acknowledgements}
 This research was supported by  the Australian Research Council Future Fellowship FT180100020,
and was funded by the Australian Government.
\end{acknowledgements}

\bibliographystyle{apsrev4-1}
%

\end{document}